\documentclass[11pt,aps,prb,superscriptaddress,showpacs]{revtex4}
\usepackage[dvips]{graphicx}
\usepackage{epsfig}
\usepackage{amsmath,amssymb,amscd}
\begin{document}
%
%
\title{
Phase diagram of S=1/2 two-leg XXZ spin ladder systems
}
%
%
\author{Keigo Hijii}
\affiliation{Department of Physics,Kyushu University, Fukuoka 812-8581, Japan}
\author{Atsuhiro Kitazawa}
\affiliation{Research Institute for Applied Mechanics, Kyushu University,
Kasuga, Fukuoka 816-858, Japan}
\author{Kiyohide Nomura}
\affiliation{Department of Physics,Kyushu University, Fukuoka 812-8581, Japan}
\date{\today}
\large
\begin{abstract}
We investigate the ground state phase diagram
of the S=1/2 two-leg $XXZ$ spin ladder system
with an isotropic interchain coupling.
In this model, there is 
the Berezinskii-Kosterlitz-Thouless transition
which occurs at the $XY$-Haldane and the $XY$-rung singlet phase boundaries.
It was difficult to determine the transition line 
using traditional methods.
We overcome this difficulty
using the level spectroscopy method combined with the twisted boundary
condition method, and we check the consistency.
We find out that the phase boundary between $XY$ phase and Haldane phase
lies on the $\Delta=0$ line. 
And we show that there exist two different $XY$ phases,
which we can distinguish investigating 
a $XX$ correlation function. 
\end{abstract}


\maketitle

\section{INTRODUCTION}
Low dimensional quantum spin systems have attracted 
some attention.
Especially, ground state properties are interesting
since the quantum fluctuation often plays the dominant role.  
For example, $S=1/2$ quantum spin ladder systems have been studied
from both theoretical and experimental points of view \cite{D-R},
related to Haldane's conjecture and high-T$_c$ superconductors.
Haldane has predicted that the properties of integer spin chains are
different from those of half-odd-integer spin chains \cite{Haldane}.
Half odd integer spin chains do not have excitation energy gaps
\cite{L-S-M}.
By contrast, for integer spin chains there is an energy gap
between the ground state and the lowest excited state\cite{Haldane}.
We can understand this fact, using the valence bond solid picture
\cite{VBS}.
From the other point of view,
Haldane's conjecture is related to spin ladder systems,
since we can describe arbitrary spin-$S$ chains as $S=1/2$ $2S$-leg
ladder systems approximately \cite{Schulz}. 

\begin{figure}
\begin{center}
\epsfig{file=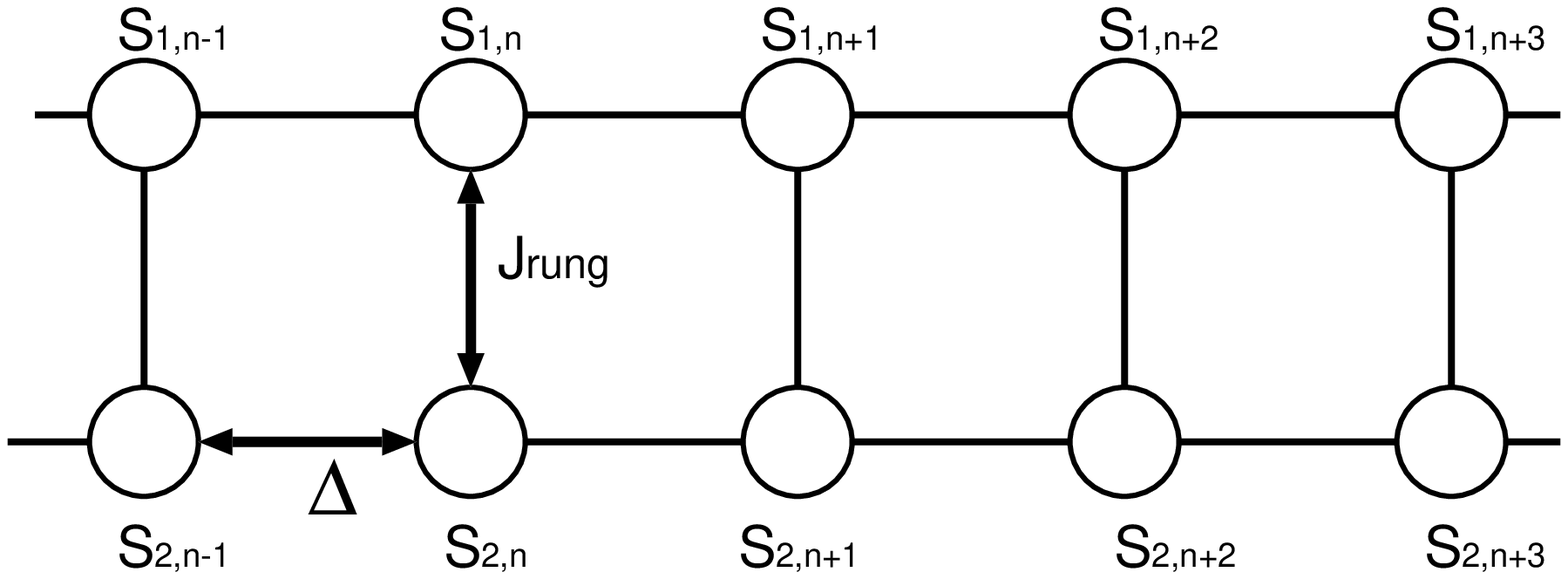,width=10cm}
\caption{Schematic structure of a S=1/2 two-leg spin ladder of Eq (1)}
\label{fig:ladder}
\end{center}
\end{figure}

The model which we study in this paper
consists of two coupled $S=1/2$ $XXZ$ chains.
We consider the following Hamiltonian,
\begin{align}
{\cal H} &= {\cal H}_{\rm leg} + {\cal H}_{\rm rung}, \nonumber \\
{\cal H}_{\rm leg} &=
\sum_{i=1}^L\sum_{\alpha=1,2} \left( 
S_{\alpha,i}^x S_{\alpha,i+1}^x + S_{\alpha,i}^y S_{\alpha,i+1}^y
+ \Delta S_{\alpha,i}^z S_{\alpha,i+1}^z
\right), \nonumber \\
{\cal H }_{\rm rung} &= J_{\rm rung} \sum_{i=1}^L {\bf S}_{1,i} 
\cdot {\bf S}_{2,i},
\label{eq:Hamiltonian}
\end{align}
where $\alpha=1,2$ are indices of chains (see Fig. \ref{fig:ladder}),
$\Delta$ is an anisotropy of the leg coupling, and $J_{\rm rung}$ is 
the rung coupling. $L$ is the system size ($L=N/2$, $N$ is the number of
sites). We shall discuss the boundary condition later.

So far, S=1/2 two-leg quantum spin ladder systems have been studied
by many authors.
Strong and Mills have presented the phase diagram using the bosonization
approach \cite{S-M}. 
Watanabe {\it et al.} have discussed the same model
using the bosonization with the aid of Wilson's renormalization
group method, and presented the phase diagram partially \cite{W-N-T}.
Narushima {\it et al.} have investigated 
Haldane-N\'eel transition \cite{N-N-T}, 
using the density matrix renormalization group (DMRG) method
\cite{DMRG}.
These theoretical studies have been done mainly with antiferromagnetic legs,
since real materials have antiferromagnetic interaction, 
and those are investigated in detail experimentally
(e.g., Sr$_{14}$Cu$_{24}$O$_{41}$).
Sr$_{14-x}$Ca${_x}$Cu$_{24}$O$_{41+\delta}$ is known as
the superconductor under high pressure\cite{U-N-A-T-M-K}.
Recently, ferromagnetic legs cases attract much interest,
for example, as Kolezhuk and Mikeska have discussed \cite{K-M}.  
There exists a linear chain with ferromagnetic interaction,
and we can expect the discovery of the ladder material 
with ferromagnetic interaction\cite{K-T-J1,K-T-J2}.
Vekua {\it et al.} have studied ferromagnetic cases using 
the bosonization approach in the weak-coupling limit \cite{V-J-M}.
They found out that an extended gapless phase appears
not only in the ferromagnetic rung coupling region but also
in the antiferromagnetic rung coupling region.
This phase was also referred by Strong and Mills \cite{S-M} too.
The stability of this gapless phase for the $\Delta <0$ region
was discussed by Legeza and Solyom. \cite{L-S}.
$\Delta= 1$ cases have been well studied using the bosonization 
\cite{S-N-T,L-O}.
In this case, the system is massive.
Lecheminant and Orignac showed that
the edge state does not exist in this model with $J_{\rm rung}>0$
under the open boundary condition (OBC) \cite{L-O}.
This means that there is no Haldane phase in $J_{\rm rung}>0$ region,
since the edge state appears in Haldane phase under the OBC.

There are several suggestions for the phase diagram of this model.
However there is no unified view.
There remain controversies in three points. 
The first point is the extent of the XY phase.
The second point is the possibility of existence of two different $XY$ phases,
and the third point is the discussion about the multicritical point 
near the Ferromagnetic phase.

For the first point, there are some previous research.
For example, Strong and Mills \cite {S-M} and Vekua {\it et al}.
\cite{V-J-M} have insisted that the $XY$ phase extend to the $\Delta >0$ region.
Watanabe {\it et al}, who used the bosonization approach,
have insisted the different extension in the $\Delta >0$ region
\cite{W-N-T}.
Narushima {\it et al}. \cite{N-N-T} and Legeza {\it et al}. 
\cite{L-S} have insisted that the XY phase does not extended 
in the $\Delta >0$ region.
They calculated the energy gap and the correlation function,using DMRG.
But Narushima {\it et al}. and Legeza {\it et al}. did not
determine the Berezinskii-Kosterlitz-Thouless (BKT) transition line,
since the correlation length becomes large compared to the system
size, especially near the $XY$ phase.
Also there are peculiar difficulties in determining the BKT transition
point \cite{LS}.

The $XY$ phase belongs to the universality class
of Tomonaga-Luttinger (TL) liquid \cite{T-L}, which is characterized by
a gapless excitation and a power-law decay of the correlation function.
In general, one component TL liquid is described by the $c=1$ 
conformal field theory (CFT) with the U(1) symmetry. 
One of the instabilities of the TL liquid is the BKT transition \cite{BKT}.
In the BKT transition, the traditional finite size scaling method
(the phenomenological renormalization group \cite{F-B})
can not be applied \cite{fsBKT,I-N}.
Moreover, since there exists the logarithmic correction,
it was difficult to determine the BKT transition point.
Recently, combining the renormalization group 
with the symmetry consideration,
the "level spectroscopy" method has been developed
in order to overcome these difficulties \cite{LS}.
Combining this method and twisted boundary condition \cite{Kitazawa,N-K},
we will numerically determine the phase boundary between the $XY$ phase
and other massive phases in this paper.

For the second point, the possibility of
two different $XY$ phases in ladder systems was pointed out
by Vekua {\it et al.}\cite{V-J-M}
in a little different model.
However, in their following paper, they did not
comment on two different $XY$ phases\cite{V-J-M2}.
We discuss this point using a different approach in this paper.
According to the Marshall-Lieb-Mattis's theorem,
the $XY1$ phase is in $J_{\rm rung}<0$ region and 
the $XY2$ phase is in the $J_{\rm rung}>0$ region 
(see section.\ref{sec:unitary}).

The third point is the discussion about the multicritical point 
near the Ferromagnetic phase.
There are different suggestions\cite{S-M,K-M,V-J-M}.
We clarified the slope toward the multicritical point at 
$\left(\Delta ,J_{\rm rung} \right)=\left(-1,0\right)$
concerning the contradicting results\cite{K-M,V-J-M}.

\begin{figure}
\begin{center}
 \epsfig{file=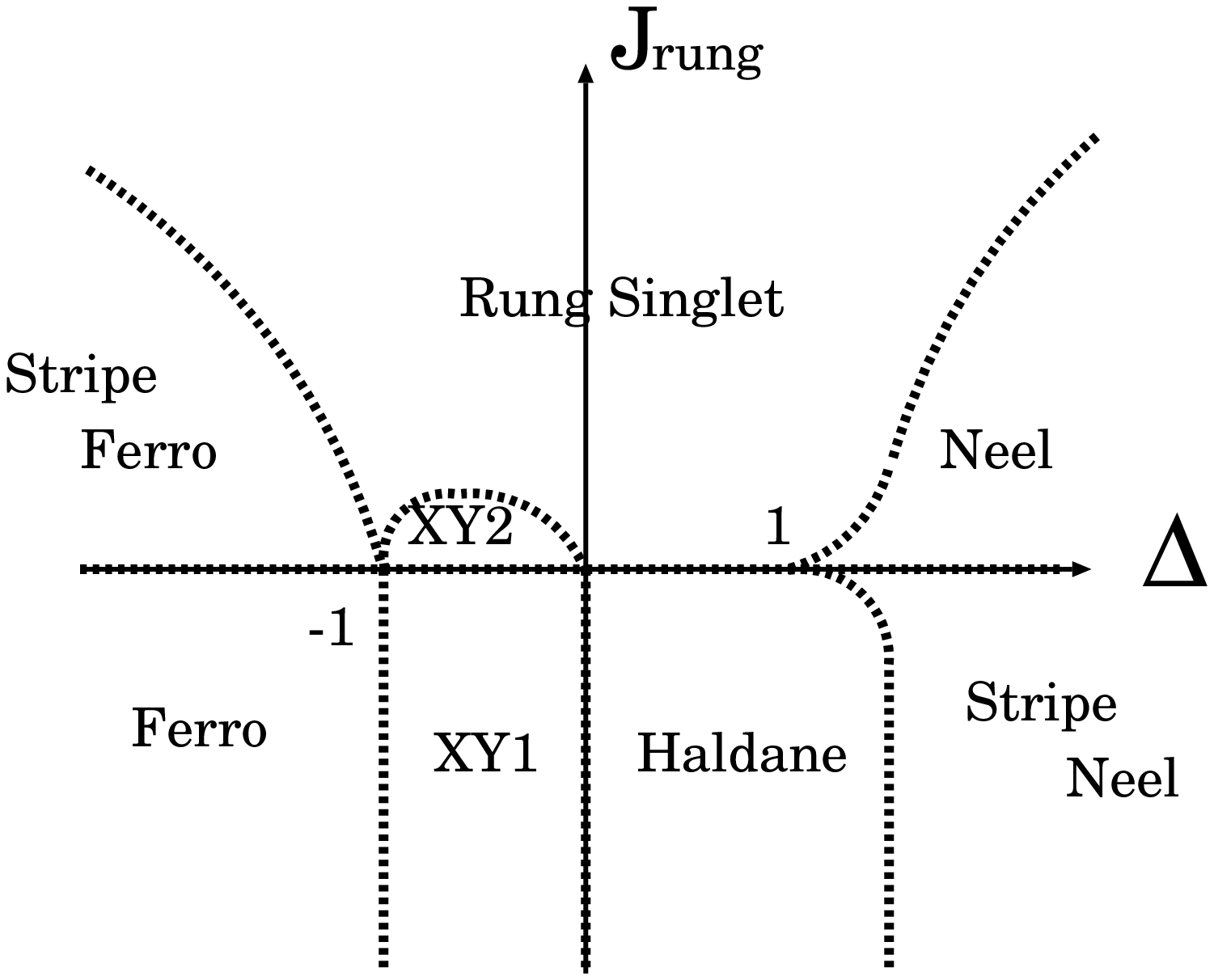,width=10cm}
\caption{
Schematic phase diagram in the S=1/2 XXZ ladder system
on $\Delta-J_{\rm rung}$ plane. Dotted lines are phase boundaries.
}
\label{fig:pdm}
\end{center}
\end{figure}

We discuss the phase diagram of S=1/2 two-leg $XXZ$ spin ladder systems
in this paper (see Fig. \ref{fig:pdm})
We organize this paper as follows.
In the next section, we consider the phase diagram 
under several limits.
In section 3, we discuss the phase boundary, 
using the variational approach.
In section 4, we discuss unitary transformations and correlation 
functions.
In section 5, we determine some phase boundaries from exact results,
considering the number of degeneracies of the ground state,
and we consider the weak coupling region.
In section 6, we numerically determine the phase boundary between the
$XY$ phase 
and the rung-singlet phase and between the $XY$ phase and the 
Haldane phase, using the level spectroscopy method,
and we discuss the multicriticality.
The last section is the conclusion.

\section{Several limits}
In this section, we consider the ground state in some limits.

In the $J_{\rm rung}=0$ case,
the system consists of two independent $S=1/2$ $XXZ$ chains,
and it was solved exactly \cite{Bethe,C-G,B-I-K}.
In the $\Delta<-1$ region, there is the ferromagnetic phase.
In the $-1< \Delta <1$ region, the XY (spin fluid) phase appears.
In the $\Delta>1$ region, the N\'eel phase appears.

In the $J_{\rm rung} \rightarrow -\infty$ case,
this model can be mapped onto an $S=1$ $XXZ$ spin chain.
The $S=1$ $XXZ$ spin chain has been studied by many authors
\cite{dN-R,C-H-S}.
In this system, there are four phases, i.e., the ferromagnetic phase, 
the $XY$ phase, the Haldane phase, and the N\'eel phase.
This Haldane phase is the gapped phase with the hidden 
$Z_2 \times Z_2$ symmetry breaking.

\begin{figure}
\begin{center}
 \epsfig{file=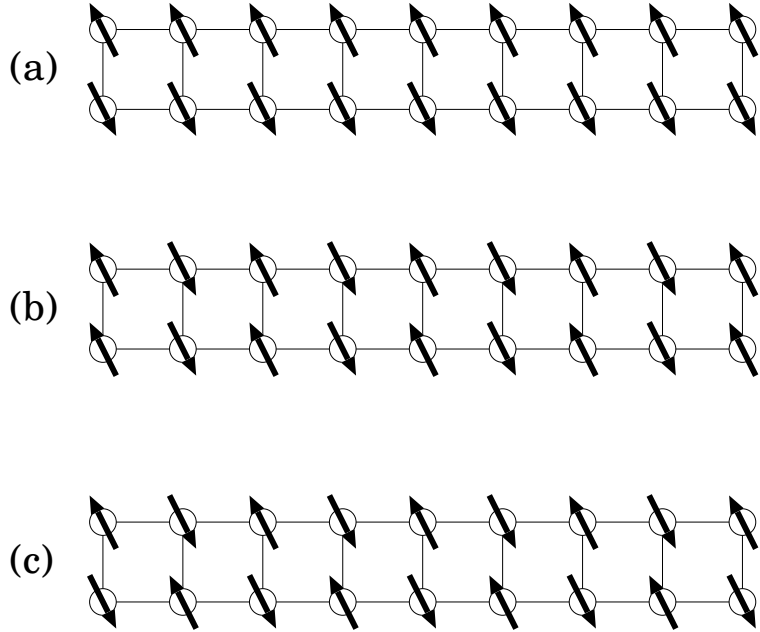,width=10cm}
\caption{
(a) Schematic picture of the Stripe Ferromagnetic state.
This phase has ferromagnetic order in the leg direction,
and antiferromagnetic order in the rung direction. 
(b) Schematic picture of Stripe N\'eel state.
This phase has antiferromagnetic order in the leg direction,
and ferromagnetic order in the rung direction.
(c) Schematic picture of N\'eel state.
This phase has antiferromagnetic order in the leg direction,
and antiferromagnetic order in the rung direction.
}
\label{fig:Stripe}
\end{center}
\end{figure}

In the $\Delta\rightarrow -\infty$ case, we can expect that
ferromagnetic ordered phases appear in each leg directions.
And we can classify them as follows.
If $J_{\rm rung}$ is positive, the ground state is called as the stripe
ferromagnetic phase \cite{V-J-M}.
The staggered magnetization appears in the rung direction.
This phase has the following correlation function:
\begin{equation}
\left< S^z_{\alpha,i}S^z_{\beta,j} \right> \propto
\left(-1\right)^{\alpha+\beta},
\end{equation}
 (see Fig. \ref{fig:Stripe}(a)).
If $J_{\rm rung}$ is negative, the ground state is the fully ferromagnetic
phase.

In the $\Delta \rightarrow + \infty$ case, we can expect that
antiferromagnetic (N\'eel) ordered phases appear in each leg direction.
If $J_{\rm rung}$ is positive, the ground state is
the N\'eel phase in the rung direction.
(see Fig. \ref{fig:Stripe}(c)).
If $J_{\rm rung}$ is negative, the ground state is 
the ferromagnetic correlation in the rung direction. 
This phase is called as the stripe N\'eel phase.
The staggered magnetization appears in the leg direction.
This phase has the following correlation function,
\begin{equation}
\left< S^z_{\alpha,i}S^z_{\beta,j} \right> \propto
\left(-1\right)^{i-j},
\end{equation}
(see Fig. \ref{fig:Stripe}(b)).

In the $\Delta \sim 0$ and $J_{\rm rung} \gg 1$ case, we can expect that 
the rung-singlet phase appears.
This state is a nondegenerate, massive phase.
We can understand this state as the direct 
product of singlets in the rung direction.

\section{Variational approach and Phase boundaries}
In this section, we roughly estimate phase boundaries using the
variational method in some limits under the periodic boundary condition
(PBC).

\subsection{Rung Singlet-Stripe Ferromagnetic phase transition}
At first, we consider the phase transition between
the rung singlet phase and the stripe ferromagnetic phase.
The variational energy of the pure rung-singlet state is given as follows:
\begin{equation}
\frac{E_{\rm RS}}{L}=
\frac{\left< \mbox{RS} \left| {\cal H}  \right| \mbox{RS} \right>}{L}
=-\frac{3}{4}J_{\rm rung},
\end{equation}
where $L$ is the system size ($L=N/2$, $N$ is the number of sites).
On the other hand,
the variational energy of the pure stripe-ferromagnetic state is given 
as follows:
\begin{equation}
\frac{E_{\rm SF}}{L}=
\frac{\left< \mbox{SF} \left| {\cal H}  \right| \mbox{SF} \right>}{L}
=\frac{1}{2}\Delta-\frac{1}{4}J_{\rm rung}.
\end{equation}
So we can roughly estimate that the phase boundary between the rung
singlet phase and the stripe ferromagnetic phase is $\Delta=-J_{\rm rung}$
in the $\Delta \rightarrow -\infty$ and
$J_{\rm rung} \rightarrow \infty$ limit.

This phase transition is related to an 
spontaneously ${\bf Z}_2$ symmetry breaking.
Therefore, this is the second order phase
transition and its universality class may be the 
two dimensional (2D) Ising type.

\subsection{N\'eel-Rung Singlet phase transition}
Next, we consider the phase transition between
the N\'eel phase and the rung singlet phase.

The variational energy of the pure N\'eel state is given as follows:
\begin{equation}
\frac{E_{\rm Neel}}{L}=
\frac{\left< \mbox{N\'eel} \left| {\cal H}  \right| \mbox{N\'eel} \right>}{L}
=-\frac{1}{2}\Delta   -\frac{1}{4} J_{\rm rung} .
\end{equation}
So we can roughly estimate the phase boundary between the rung singlet 
phase and the N\'eel phase is $\Delta=J_{\rm rung}$,
in $\Delta \rightarrow \infty$ and
$J_{\rm rung} \rightarrow \infty$ limit.
This phase transition is related to the ${\bf Z}_2$ symmetry breaking.
Therefore, this phase transition is the second order phase
transition and its universality class may be the 2D-Ising type.

\section{Unitary transformation and Phase transition}
\label{sec:unitary}
In this section, we treat the bipartite systems.
That is,
the system size ($L=N/2$) is even under the periodic boundary condition,
or the system size ($L$) is arbitrary under the open boundary
condition, where $N$ is the number of spins.

\subsection{$XY$ phase - Ferromagnetic phase transition}
\begin{figure}
\begin{center}
\epsfig{file=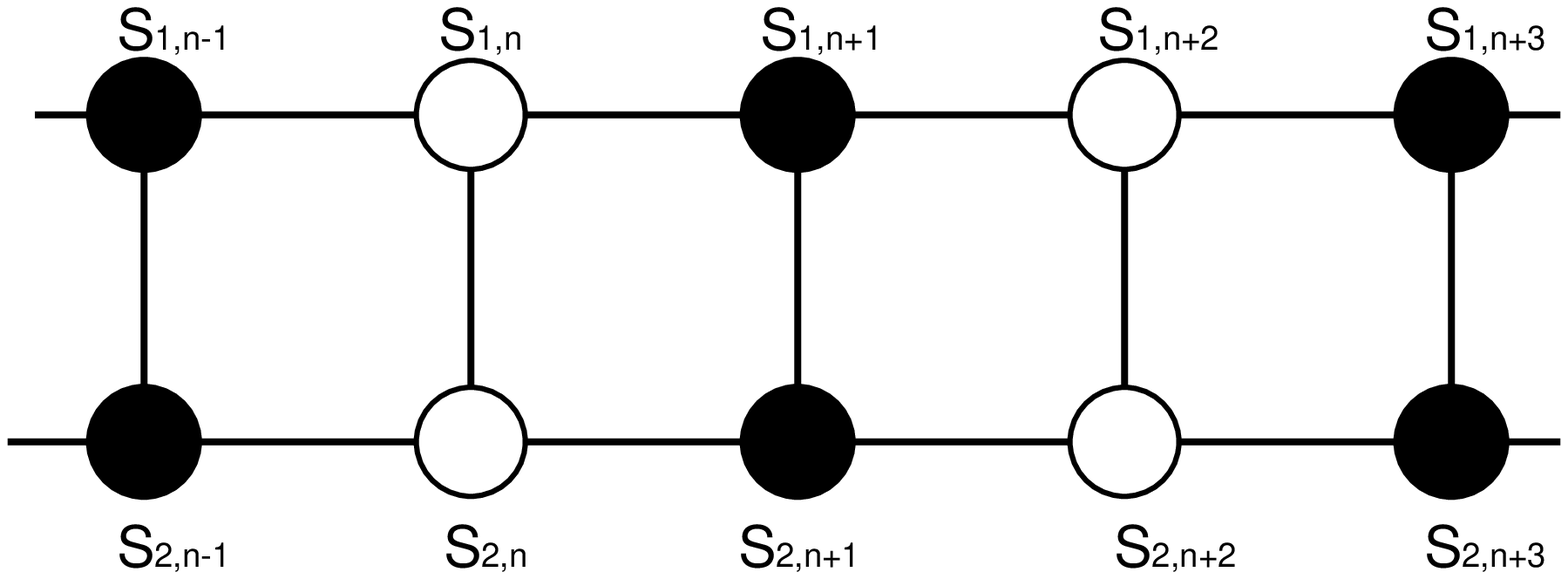,width=10cm}
\caption{
Schematic figure for the unitary transformation in the $J_{\rm rung}<0$
case.
Empty circles are uncchanged.
Full circles are $S^{\pm}_{\alpha,j} \rightarrow -S^{\pm}_{\alpha,j}$
}
\label{fig:nrung}
\end{center}
\end{figure}

In the ferromagnetic coupling case ($\Delta<-1,J_{\rm rung}<0$),
we found that the ferromagnetic phase appears.
Vekua {\it et al.} have obtained that the phase boundary 
between $XY$ phase and the ferromagnetic phase 
locates at $\Delta=-1$ using the instability analysis of the
spin wave theory for the ferromagnetic phase \cite{V-J-M}.
Here we try to use another approach for this phase transition.
We transform the original Hamiltonian (\ref{eq:Hamiltonian})
using a following unitary operator:
\begin{equation}
U=\exp \left[ i\pi \sum_{j,\alpha}S^z_{\alpha,2j+1}  \right]
\label{eq:unitary1}
\end{equation}
(see Fig \ref{fig:nrung}).
This unitary operator transform 
$S^{\pm}_{\alpha,2j+1} \rightarrow -S^{\pm}_{\alpha,2j+1}$.
Especially, in the $\Delta=-1$ case,
the transformed Hamiltonian is described as the pure Heisenberg model
with a negative coupling constant.
Now we can choose a set of basis vectors as eigenvectors which
diagonalize $S^z$.
Then all off-diagonal elements of the transformed Hamiltonian 
have a negative sign.
From Perron-Frobenius's theorem, 
the ground state is unique in the fixed 
$S^z_T=\sum_{j,\alpha}S^z_{\alpha,j}$ space in the finite system,
and we can choose that all the coefficients are positive.
Since the transformed system has an $SU(2)$ symmetry 
in the $\Delta=-1$ case, 
states with $S^z_T=0,\pm1,\pm2,\cdots$ are degenerate (SU(2) Ferro).
This means that $J_{\rm rung}<0, \Delta=-1$ line is the phase boundary,
between the fully ferromagnetic phase ($\Delta<-1$) 
and the $XY$ phase ($\Delta>-1$). 

Especially, in the $(J_{\rm rung}=0,\Delta=-1)$ case 
and the system size $L(=N/2))$, where $L$ is even,
there are the number of $(L+1)^2$ degenerated ground states,
since this system has an $SU(2) \times SU(2)$ symmetry.
We can consider that this point is the multicritical point,
where the BKT transition line meets the 2D Ising type phase transition
line.

\subsection{Off critical case.1 ($J_{\rm rung}<0$)}
In this case, we consider the unitary transformation (\ref{eq:unitary1}),
(see Fig \ref{fig:nrung}).
After this unitary transformation, all off-diagonal elements of
Hamiltonian become negative.
From the discussion of the previous subsection,
signs of correlation functions are 
represented in the original Hamiltonian (\ref{eq:Hamiltonian}) as follows,
\begin{equation}
\left(-1\right)^{i-j}
\left<S^x_{\alpha,i}S^x_{\beta,j}\right> >0.
\end{equation}
This corresponds to the Ferromagnetic phase, the $XY1$ phase,
Haldane phase and the Stripe N\'eel phase.

\subsection{Off critical case.2 ($J_{\rm rung}>0$)}
\begin{figure}
\begin{center}
 \epsfig{file=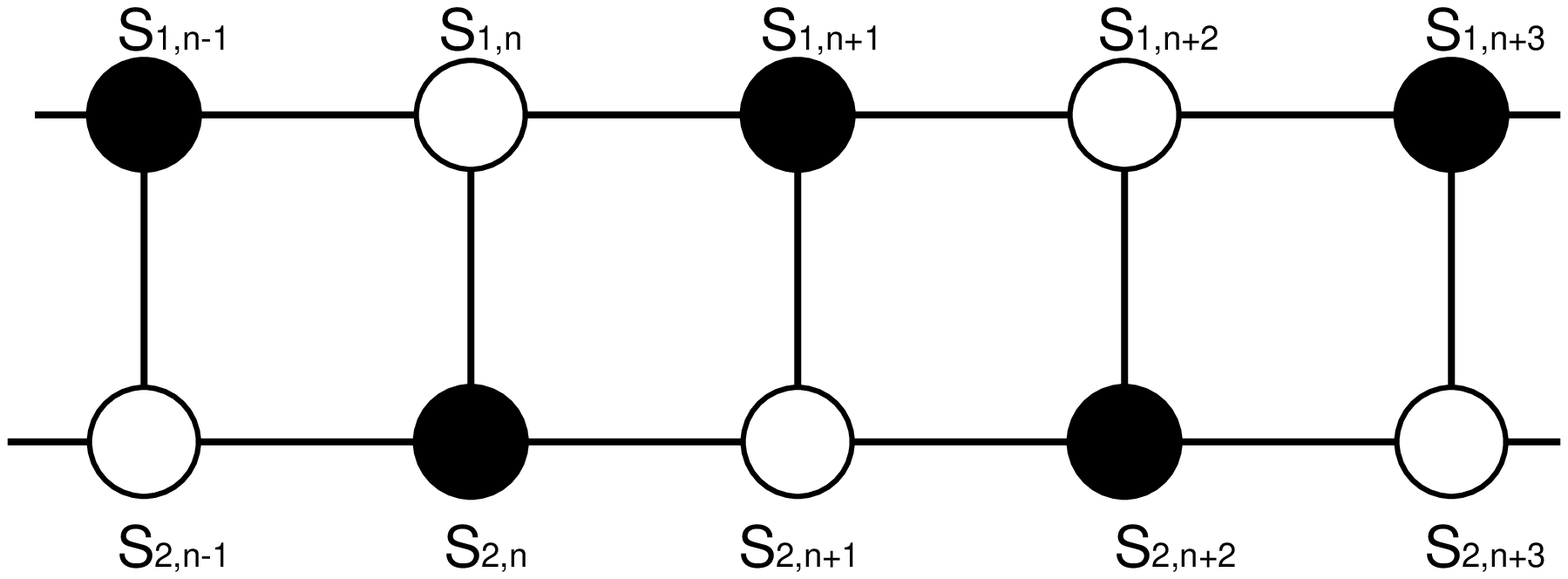,width=10cm}
\caption{
Schematic figure for the unitary transformation in the $J_{\rm rung}>0$
case.
Empty circles are uncchanged.
Full circles are $S^{\pm}_{\alpha,j} \rightarrow -S^{\pm}_{\alpha,j}$
}
\label{fig:prung}
\end{center}
\end{figure}

In this case, we consider the following unitary transformation:
\begin{equation}
 U=\exp \left[ i\pi \sum_{j} \left( S^z_{1,2j+1} + S^z_{2,2j} \right)
	\right]
\label{eq:unitary2}
\end{equation}
(see Fig \ref{fig:prung}).
This unitary operator transform 
$S^{\pm}_{1,2j+1}\rightarrow -S^{\pm}_{1,2j+1}$ and 
$S^{\pm}_{2,2j}\rightarrow -S^{\pm}_{2,2j}$.
After this unitary transformation, all off-diagonal elements of
Hamiltonian become negative.
From the Marshall-Lieb-Mattis's theorem\cite{M-L-M},
this case has following signs of the correlation function 
in the original Hamiltonian (\ref{eq:Hamiltonian}):
\begin{equation}
\left(-1\right)^{\alpha+\beta}\left(-1\right)^{i-j}
\left<S^x_{\alpha,i}S^x_{\beta,j}\right>>0
\end{equation}
This corresponds to the Stripe Ferromagnetic phase, the $XY2$ phase,
the rung singlet phase and N\'eel phase.

\subsection{Off critical case.3 ($\Delta=-1,J_{\rm rung}>0$)}
In this subsection, we consider ($\Delta=-1,J_{\rm rung}>0)$ case.
Using the unitary transformation (\ref{eq:unitary1}),
we can see that the system has an $SU(2)$ symmetry.
From Mermin-Wagner's theorem, since there is no long range order
with the spontaneously continuous symmetry breaking,
the ground state is not the Stripe Ferromagnetic phase.
Since there is no soft mode with the wave number $q=\pi$
for the transformed Hamiltonian,
there is no possibility that it is described as
a $k=1,2$ $SU(2)$ Wess-Zumino-Witten model, which is massless.

From the Marshall-Lieb-Mattis's theorem \cite{M-L-M}, 
the sign of the correlation function
transformed by unitary operator (\ref{eq:unitary1}), 
\begin{equation}
\left< S^{x}_{1,i}S^{x}_{2,j}\right><0.
\end{equation}
Thus this ground state is not an $SU(2)$ ferromagnetic phase.

Therefore we can conclude that the ground state is the rung singlet
phase in the ($\Delta=-1,J_{\rm rung}>0$) case.

\section{Approach from decoupled chains}
In this section, we discuss the phase diagram from the exact solutions.
This system with $J_{\rm rung}=0$ consists of two independent chains.
From the exact solution we know that $S=1/2$ $XXZ$ chain with anisotropy
$\left|\Delta\right|>1$ is massive, and the ground state is twofold
degenerate. 
Now the system is two decoupled chains, therefore, the ground state 
is fourfold degenerate.
In following subsections, we discuss the neighborhood of 
$J_{\rm rung}=0$.

\subsection{Ferromagnetic phase - Stripe Ferromagnetic phase transition}
In $\Delta<-1$ region, we can find two different two ferromagnetic
ordered phases, as suggested by Vekua {\it et al} \cite{V-J-M}.
The usual ferromagnetic ordered phase is two fold degenerate.
However on the $\Delta<-1,J_{\rm rung}=0$ line, the ground state
is four fold degenerate.
In the first case all spins of both chains are up; 
in the second case all spins of both chains are down;
in the third case all spins of 1-chain is up and all spins of 2-chain 
is down;
and in the forth case all spins of 1-chain is down and 
all spins of 2-chain
is down (see Fig.(\ref{fig:Stripe})(a)).
These four states have the same energy exactly on the $J_{\rm rung}=0$.
A nonzero $J_{\rm rung}$ term breaks this degeneracy.
This means that $\Delta<-1,J_{\rm rung}=0$ line is the phase boundary.
In addition, 
the system, which consists of two independent gapped system is gapped.
So this phase transition is the first order transition.

\subsection{N\'eel phase - Stripe N\'eel phase transition}
The discussion in the previous subsection can be applied to
the antiferromagnetic ordered phase similarly.

Firstly, we consider the single $S=1/2$ $XXZ$ chain.
From the exact solution, in $\Delta > 1$ region,
the ground state is the antiferromagnetic
(N\'eel) ordered state, 
which is two fold degenerate. 
Secondly, we consider independent double $S=1/2$ $XXZ$ chains.
The ground state is fourfold degenerate
in $\Delta > 1$ region,
since this total system is described as the direct product of 
the single chain system.
These ground states are shown in  Fig.\ref{fig:Stripe} (b),(c).
Lastly, we consider the coupled $S=1/2$ $XXZ$ chains
described by the Hamiltonian (\ref{eq:Hamiltonian}).
The rung-coupling term breaks the four fold degeneracy in the above case.
In $J_{\rm rung}<0$ case,
the ferromagnetic ordered state in the rung direction
(see, Fig.\ref{fig:Stripe} (b)), 
which is called as the stripe N\'eel phase.
In $J_{\rm rung}>0$ case,
the antiferromagnetic ordered state in the rung direction
(see, Fig.\ref{fig:Stripe} (c)), 
which is called as the N\'eel phase.
Both states are antiferromagnetic ordered in the leg direction.

Since the system is gapped in the $J_{\rm rung}=0$ case,
this phase transition between the N\'eel phase and the stripe N\'eel phase
is the first order transition.

\subsection{Weak coupling region}
In this subsection, we consider the weak coupling S=1/2 XXZ chain,
$-1 \le \Delta \le 1$ and $\left|J_{\rm rung}\right| \ll 1$.

At first, we consider two independent chains, $J_{\rm rung}=0$ case.
This case is exactly solved using Bethe ansatz\cite{Bethe,C-G,B-I-K}.
One chain in the $-1<\Delta<1$ region is described by TL liquid
\cite{T-L}.
Two independent chains are described by the direct product of
two TL liquids.
This model has critical properties in extended regions in the
parameter space.
Here we consider the case that there are some perturbations for this
system. 

These systems can be analyzed using the bosonization method
\cite{S-M,V-J-M}.
Then two decoupled chains are described in two bosonic fields,
$(\phi_1,\theta_1)$ and $(\phi_2,\theta_2)$ of each other.
We obey the notation used in Ref.\cite{V-J-M} and introduce 
the symmetric and antisymmetric combination of the bosonic field,
\begin{equation}
\left\{
\begin{array}{c}
\phi_\pm \propto \phi_1 \pm \phi_2, \\
\theta_\pm \propto \theta_1 \pm \theta_2. \\
\end{array}
\right.
\end{equation}
We consider the case with $J_{\rm rung}\neq 0$. 
Behavior of the symmetric field
is
governed by the effective sine-Goldon model.
$\cos\sqrt{\frac{2\pi}{K_-}}\theta_-$ is always relevant. 
$\cos\sqrt{2\pi K_+}\phi_+$ is relevant for $\Delta>0$, 
irrelevant for $\Delta<0$,
where $K_{\pm}$ are constant \cite{V-J-M}.

The phase transition between the $XY$ phase and 
the nondegenerate massive phase is the $K=4$ BKT type transition 
when the system has a simple $U(1)$ symmetry, 
where $K$ is the TL parameter.
In this case, a $(S^z_T=0,q=\pi)$ mode is always massive,
and a $(S^z_T=0,q=0)$ mode becomes massless.

We can distinguish two different $XY$ phases.
One $XY$ phase ($XY1$) is in $J_{\rm rung}<0$, 
another $XY$ phase ($XY2$) is in $J_{\rm rung}>0$ region. 
In the $J_{\rm rung}>0$ region,
from the Marshall-Lieb-Mattis's theorem,
$\left<S^x_{1,j}S^x_{2,j}\right> < 0$.
On the other hand,
in the $J_{\rm rung}<0$ region, $\left<S^x_{1,j}S^x_{2,j}\right> > 0$.
Therefore these two $XY$ phases have different symmetry.
A $J_{\rm rung}=0$ line is the second order phase transition line.
The system on the $J_{\rm rung}=0$ line is described as two component 
TL liquid. 
This second order phase transition is explained as 
that one component of the TL liquid remains massless, 
but another component become massive.

Here we summarize this subsection in the CFT language.
In the $\left|\Delta\right|<1$ region, 
the system is described by the central charge 
$c=2$ CFT on the $J_{\rm rung}=0$ line.
And there exists a region which is described as $c=1$ CFT
in the $0<\left|J_{\rm rung}\right| \ll 1$ and $-1 \leq\Delta\leq 0$.

\section{Phase Transitions and Numerical Results}
In this section, we determine the critical points and the universality class.
We show the numerical results of $N=12,16,20,24$ systems using
the exact diagonalization, and the level spectroscopy method with the twisted
boundary condition \cite{LS,Kitazawa,N-K}.

\subsection{XY-Rung singlet transition}
\begin{figure}
\begin{center}
\epsfig{file=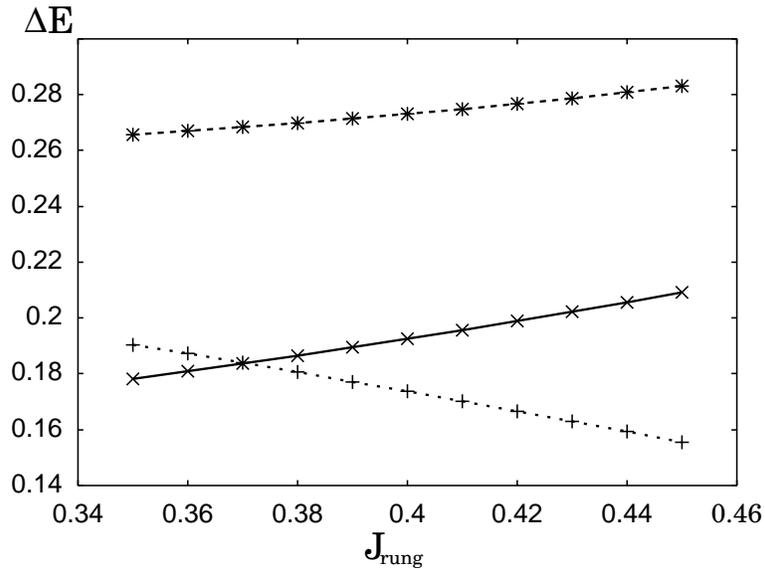,width=10cm}
\caption{
Excitation energies of $L=12(=N/2)$, $\Delta=-0.5$ near the XY-Rung
 Singlet transition point. $\times$'s are $S^z_T=\pm2$,$q=0$,$P=1$ under
the periodic boundary condition. $+$'s are $S^z_T=0$,$P^*=1$ under
the twisted boundary condition.$*$'s are $S^z_T=0$,$P^*=-1$ under
the twisted boundary condition.
 }
\label{fig:cp-xy-rs}
\end{center}
\end{figure}

We can expect that this phase transition is the BKT type.
Using the level spectroscopy method, 
we determine critical points of this phase transition.
Considering symmetries, we can identify critical points as cross points
of two low-lying excitations.
One has quantum numbers $S^z_T=\pm 2, P=1, q=0$ 
under the periodic boundary condition (PBC),
another has quantum numbers $S^z_T=0, P^*=1$ under the
twisted boundary condition (TBC)
$(S^{x,y}_{i,j} = - S^{x,y}_{i,j+L})$.
Here $S^z_T$ is total magnetizations of the system 
($S^z_T\equiv\sum_{i,\alpha}S^z_{\alpha,i}$),
and $P$ is the parity defined under the following transformation 
$(S^{\alpha}_{i,j} \leftrightarrow S^{\alpha}_{i,L-j+1})$,
$P^*$ is the parity under the TBC.
$q$ is the wave number defined under the periodic boundary condition,
$L$ is the system size which is a half of the number of sites.
We show crossing points in Fig. \ref{fig:cp-xy-rs}.

\begin{figure}
\begin{center}
\epsfig{file=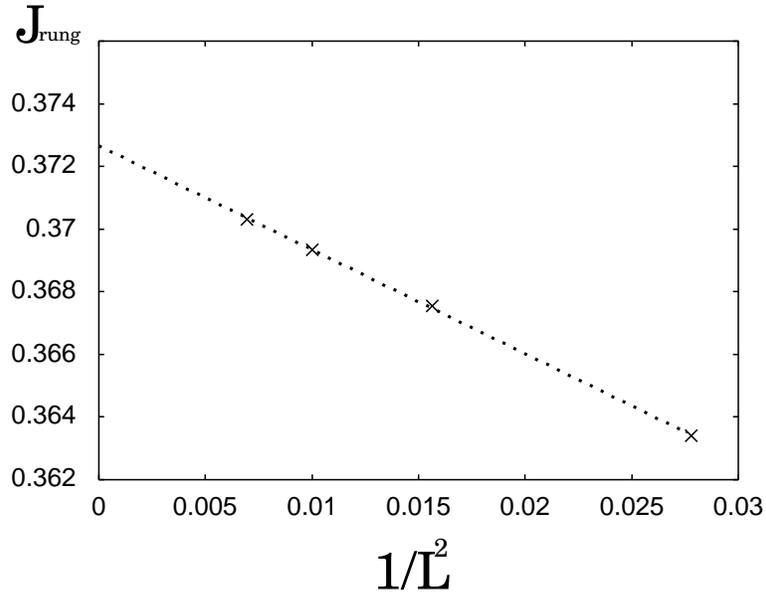,width=10cm}
\caption{Size dependence of the crossing points. The extrapolated 
value is 0.3727 at $\Delta=-0.5$.
Dotted line is determined using a linear least method.
}
\label{fig:sizeeff}
\end{center}
\end{figure}

\begin{figure}
\begin{center}
\epsfig{file=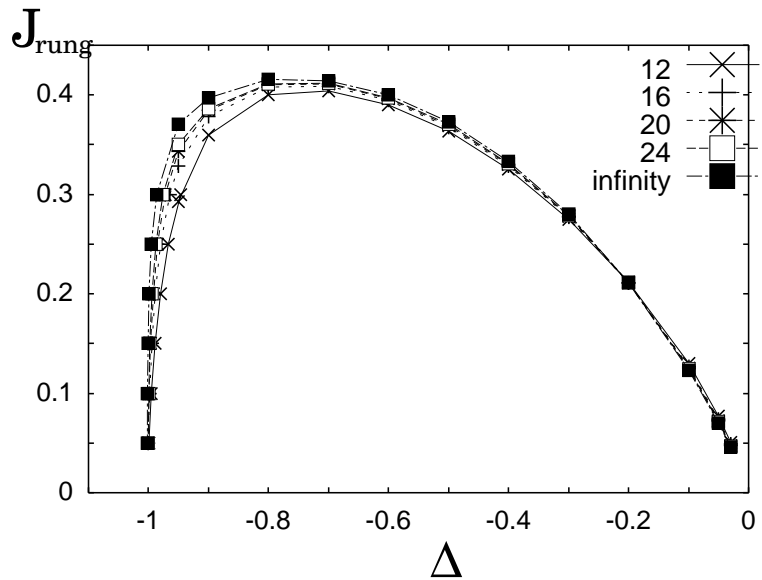,width=10cm}
\caption{
crossing points and their size dependence for XY-Rung Singlet transition,
for N=12,16,20,24,and extrapolated values.
 }
\label{fig:pd-xy-rs}
\end{center}
\end{figure}

However values of crossing points have size dependence.
We can remove logarithmic corrections, which is proportional to
$1/\ln L$, using the level spectroscopy method.
However, there remain other corrections 
which come from the lattice structure.
For example, there are correction terms 
from the $x=4$ irrelevant field ($L_{-2}\overline{L}_{-2}{\bf 1}$)
\cite{Cardy}.
Therefore we extrapolate crossing points as follows,
\begin{equation}
J_{\rm rung}^{\rm cross} \left( L \right)=J_{\rm rung}^{\rm cross} 
\left( \infty \right)
 + a \frac{1}{L^2} + o \left( \frac{1}{L^4} \right)
\end{equation}
or
\begin{equation}
\Delta^{\rm cross} \left( L \right)=\Delta ^{\rm cross} 
\left( \infty \right) + a
 \frac{1}{L^2} + o \left( \frac{1}{L^4} \right)
\end{equation}
where $a$ is a fitting parameter. We neglect higher order terms 
$o\left(\frac{1}{L^4} \right)$.
In Fig. \ref{fig:sizeeff}, we show the size dependence of the crossing 
points and its extrapolation.
Then we can find the phase boundary under the infinite system limit
in Fig.(\ref{fig:pd-xy-rs}).

In order to confirm the consistency of our results,
we need check the universality class using CFT \cite{CFT}.
In this case, the critical theory which describes the transition line 
is the $c=1$ CFT.
The central charge appears as an universal finite size correction
for the ground state energy under the PBC
\cite{B-C-N,Affleck,Cardy1},
\begin{equation}
E_{\rm g} \left( L \right)=\epsilon L -\frac{\pi v }{6L}
\left( c +o \left( \frac{1}{ \left( \ln L \right)^3 }\right)\right)
\end{equation}
where $L$ is the system size, $\epsilon$ is the ground state energy per
site in the infinite system, $v$ is the spin wave velocity.
Now $o\left( \frac{1}{ \left( \ln L \right)^3 }\right)$ correction
terms are small enough for numerical data, so we neglect it.
In order to numerically determine the central charge,
we should obtain $v$ in addition to the ground state energy.
We can obtain the spin wave velocity as follows:
\begin{equation}
v \left( L \right) = \frac{L}{2\pi} \left(
E \left( q=\frac{2\pi}{L} \right) - E_g
\right).
\end{equation}
where $q$ is a wave number. 
Then we extrapolate $v \left( L \right)$ as
\begin{equation}
v\left( L \right) = v \left( \infty \right) + a \frac{1}{L^2} 
+b \frac{1}{L^4} + \mbox{higher order},
\end{equation}
where $a$ and $b$ are fitting parameters. We neglect higher order terms.

\begin{figure}
\begin{center}
\epsfig{file=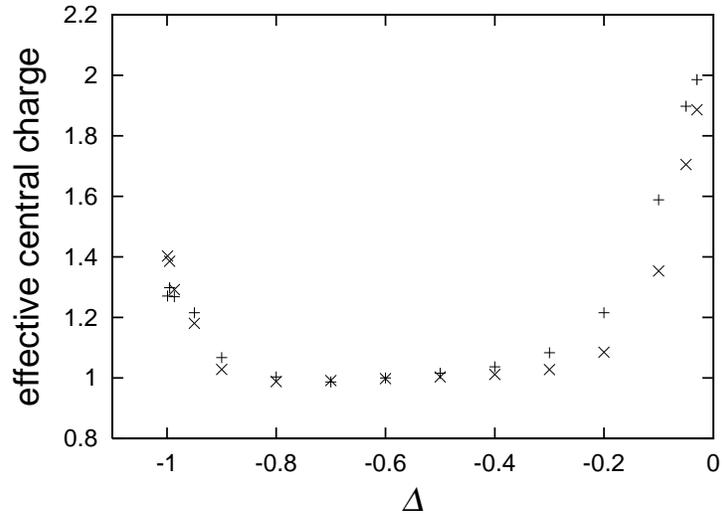,width=10cm}
\caption{The effective central charge on XY-Rung Singlet transition
 line. $\times$'s are extrapolated using $L(=N/2)=6,8,10$.
$+$'s are extrapolated using $L(=N/2)=8,10,12$.
}
\label{fig:cc-xy-rs}
\end{center}
\end{figure}

\begin{figure}
\begin{center}
 \epsfig{file=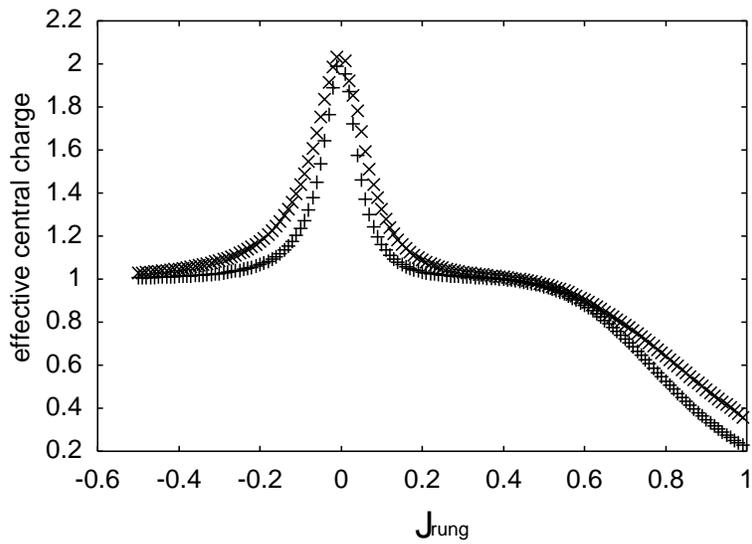,width=10cm}
\caption{
The effective central charge $\tilde{c}$ 
as a function of $J_{\rm rung}$ at $\Delta=-0.5$.
$\times$ are extrapolated using $N=12,16,20$.
$+$ are extrapolated using $N=16,20,24$.
}
\label{fig:ecc}
\end{center}
\end{figure}

Here we define the effective central charge in numerical calculations 
as follows,
\begin{equation}
E_{\rm g} \left( L \right)=\epsilon L -\frac{\pi v \tilde{c}}{6L},
\end{equation}
where $\tilde{c}$ is the effective central charge.
The effective central charge is equivalent to the central charge 
on the critical point or in the TL liquid phase.
And this changes rapidly from $\tilde{c}=1$ (the TL phase)
to $\tilde{c}\rightarrow0$ (the massive phase) \cite{I-N}.
In Fig.(\ref{fig:cc-xy-rs}), we show the effective central charge 
obtained on the transition points.
In this figure, we find that the central charge 
decreases from $\tilde{c}=2$ ($J_{\rm rung}=0$) 
to $\tilde{c}=1$ (far from $J_{\rm rung}=0$).
This reflects Zamolodchikov's c-theorem \cite{c-theo}.
And we consider the phase transition between the rung-singlet phase
(massive )and the $XY$ phase (massless).
We see that the effective central charge $\tilde{c}$ 
is $1$ in normal $XY$ phases, 
and $\tilde{c}$ is $2$ in $XY$ phase in two independent chains
$J_{\rm rung}=0$, 
and $\tilde{c}$ becomes zero increasing the system size
($L\rightarrow\infty$) in the rung-singlet phase
(see Fig.(\ref{fig:ecc})).

\begin{figure}
\begin{center}
\epsfig{file=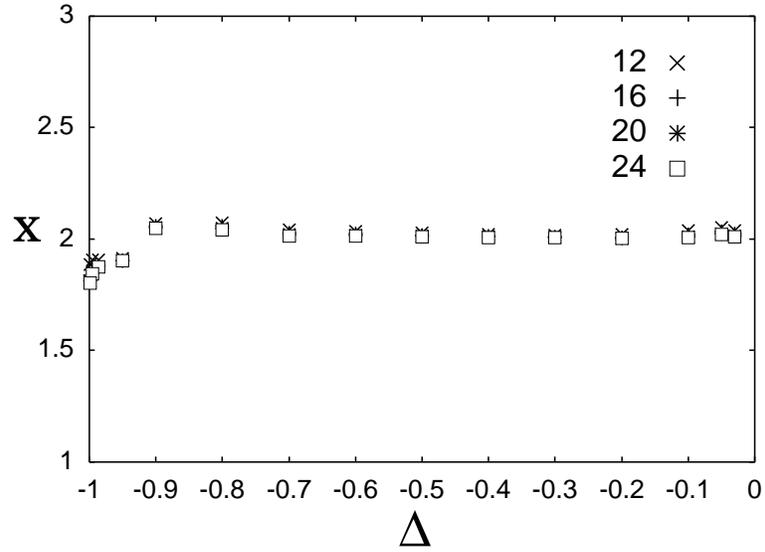,width=10cm}
\caption{Size dependence of scaling dimensions removed logarithmic
 correction on the XY-Rung Singlet transition line
 for the system size $N=(2L)=12,16,20,24$.
 }
\label{fig:sd-xy-rs}
\end{center}
\end{figure}

And then we should calculate scaling dimensions removing logarithmic
corrections.
From CFT, scaling dimensions in the finite size system under the PBC
are related to excitation energies as follows \cite{Cardy2}:
\begin{equation}
 E_i \left( L \right) - E_g \left( L \right)= \frac{2\pi v x_i}{L},
\end{equation}
where $E_{i}$ is an excitation energy, $v$ is the spin wave velocity,
$L$ is the system size, $x_i$ is the scaling dimension.
In fact, there exists additional logarithmic corrections.
\begin{table}
\caption{
$P=1^*$ is even parity under TBC,
$P=-1^*$ is odd parity under TBC,
, $q$ is the wave number, $x$ is the scaling dimension,
$y_l \propto 1/\ln L$
}
\label{tb:sdim}
\begin{center}
\begin{tabular}{|c|c|c|c|c|c|}
\hline
$S_z^T$ & P & q & BC & x &  abbr \\ 
\hline
$\pm2$ & 1 & 0 & PBC & $\frac{1}{2}-\frac{y_1}{4}$  & $x_{\pm2,0}$  \\  
\hline
0 & -1$^*$ &  & TBC & $\frac{1}{2}+\frac{y_1}{4}-\frac{y_2}{2}$ &  $x_{0,sin}^{TBC}$  \\  
\hline
0 & 1$^*$ &  & TBC & $\frac{1}{2}+\frac{y_1}{4}+\frac{y_2}{2}$ &  $x_{0,cos}^{TBC}$  \\  
\hline
$\pm4$ & 1 & 0 & PBC & $2-y_1$ &  $x_{\pm4,0}$  \\  
\hline
0 & 1 & 0 & PBC & $2-y_1 \left( 1+\frac{4t}{3} \right)$  & $x_{marg}$  \\  
\hline
0 & -1 & 0 & PBC & $2+y_1 $ &  $x_{0,sin}$  \\  
\hline
0 & 1 & 0 & PBC & $2+2y_1 \left( 1+ \frac{2t}{3} \right) $ &  $x_{0,cos}$  \\  
\hline
\end{tabular} 
\end{center}
\end{table}
From table (\ref{tb:sdim}), we obtain the following relation:
\begin{equation}
2 x_{\pm2,0} + x_{0,{\rm sin}}^{\rm TBC} + x_{0,{\rm cos}}^{\rm TBC} =2,
\end{equation}
removing leading logarithmic corrections.
In Fig. \ref{fig:sd-xy-rs}, we show its size dependence.

Further more, we check the consistency of BKT phase transition.
We calculate the ratio of two excitation energies, $S^z_T=1$ and $S^z_T=2$.
\begin{eqnarray}
\Delta E_{1}&=&E(S^z_T=1) - E_{\rm g}, \nonumber \\
\Delta E_{2}&=&E(S^z_T=2) - E_{\rm g},\nonumber \\
f&=&\frac{\Delta E_2}{\Delta E_1},
\label{eq:ratio}
\end{eqnarray}
where $E_{\rm g}$ is the ground state energy, $E(S^z_T=1,2)$ is the lowest 
energy with the quantum number $S^z_T=1,2$.
This $f$ is equivalent to the ratio of scaling dimensions
on the TL liquid.
For Gaussian model \cite{K-B}, scaling dimensions are described as
\begin{equation}
x_{m,n}=\frac{1}{2}\left( \frac{m^2}{K} + n^2K \right),
\label{eq:scalingdimension}
\end{equation}
under the PBC, where $m$ is related to the magnetization of the system,
$n$ is not related to any conserved quantities,
$K$ is the TL parameter. $m$ and $n$ are integers.
From eq.(\ref{eq:ratio}) and eq.(\ref{eq:scalingdimension})
$f$ should be $4$ in one component TL liquid phase.

However, when the system consists of two independent chain,
this $f$ should be 2 in TL liquid phase for the following reason.
In $S^z=1$ case, one chain has the ground state energy $E_{\rm g}$,
another chain has the excitation energy $E(S^z=1)$,
thus the system has the excitation energy $\Delta E_1$.
In $S^z=2$ case, each chain has the excitation energy $E(S^z=1)$,
so the system has the excitation energy $2\Delta E_1$
(see Appendix \ref{sec:ind-chain}).
In this case, $f$ should be 2.

\begin{figure}
\begin{center}
 \epsfig{file=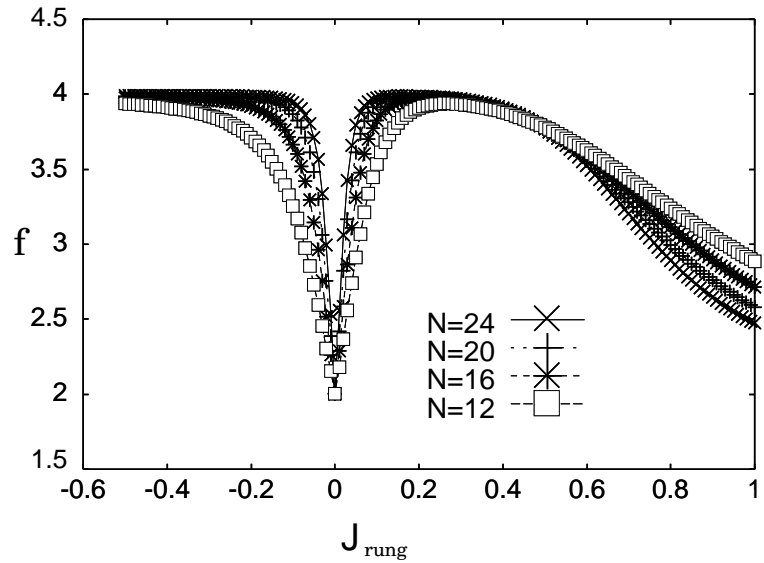,width=10cm}
\caption{
The ratio of energy gaps as a function of $J_{\rm rung}$
at $\Delta=-0.5$ and its size dependence.
}
\label{fig:ratio}
\end{center}
\end{figure}

On the other hand, in a massive phase, the energy of two independent
magnons is twice the energy of one magnon in $L\rightarrow\infty$ limit.
If the interaction of magnons is repulsive, 
then the ratio is more than 2 in the finite system.
If the interaction of magnons is attractive then the ratio is less than 2
in the finite system.
We show the ratio $f$ in Fig. \ref{fig:ratio}.
This figure support the second order phase transition between
the $XY1$ phase and $XY2$ phase, together with Fig. \ref{fig:ecc}.

\subsection{$XY$-Haldane transition}
\begin{figure}
\begin{center}
\epsfig{file=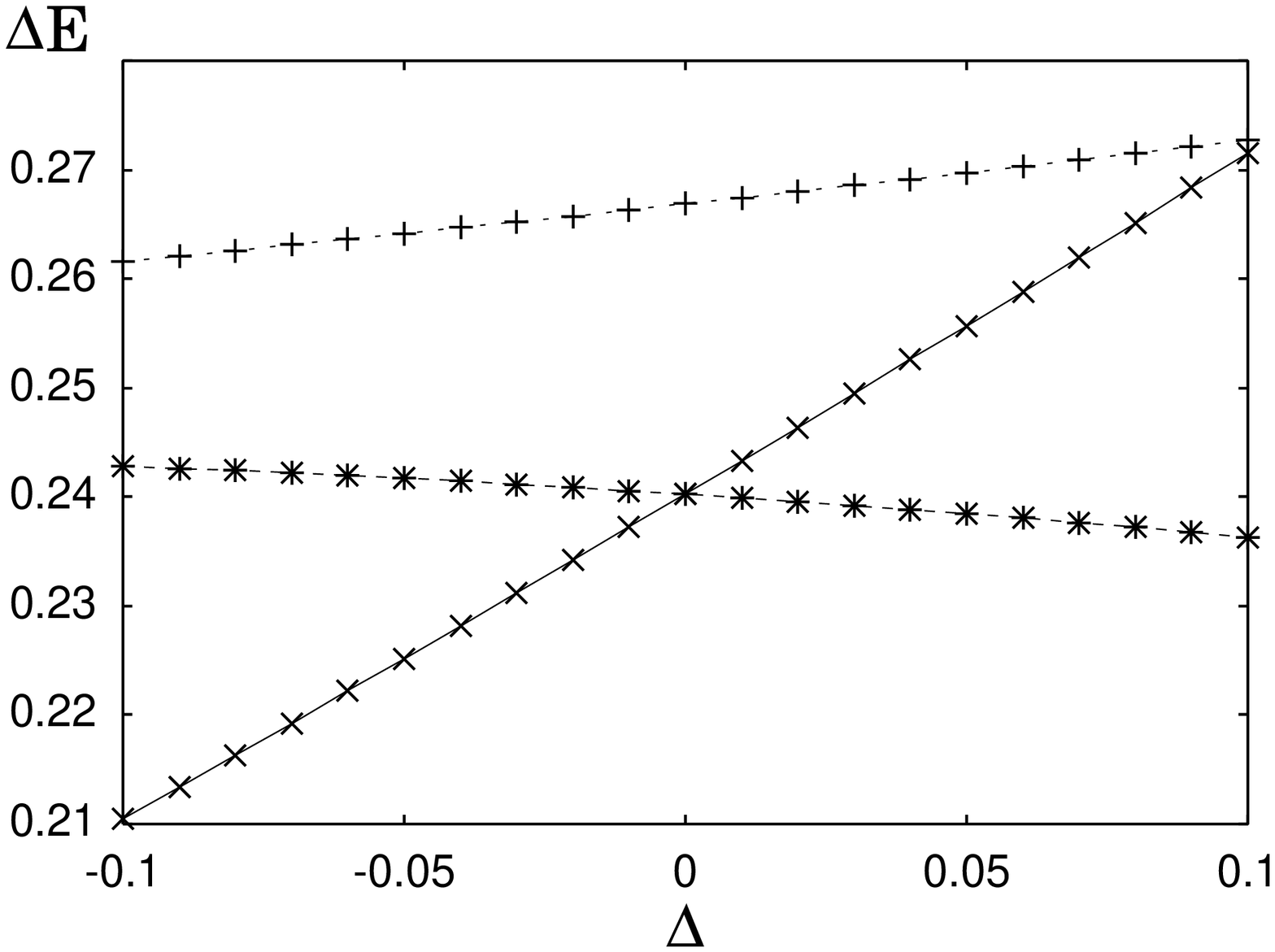,width=10cm}
\caption{
Excitation energies of $L=12(=N/2)$, $J_{\rm rung}=-0.5$ near the XY-Haldane
transition point. $\times$'s are $S^z_T=\pm2$,$q=0$,$P=1$ under
the periodic boundary condition. $+$'s are $S^z_T=0$,$P^*=1$ under
the twisted boundary condition.$*$'s are $S^z_T=0$,$P^*=-1$ under
the twisted boundary condition.
 }
\label{fig:cp-xy-haldane}
\end{center}
\end{figure}

\begin{figure}
\begin{center}
\epsfig{file=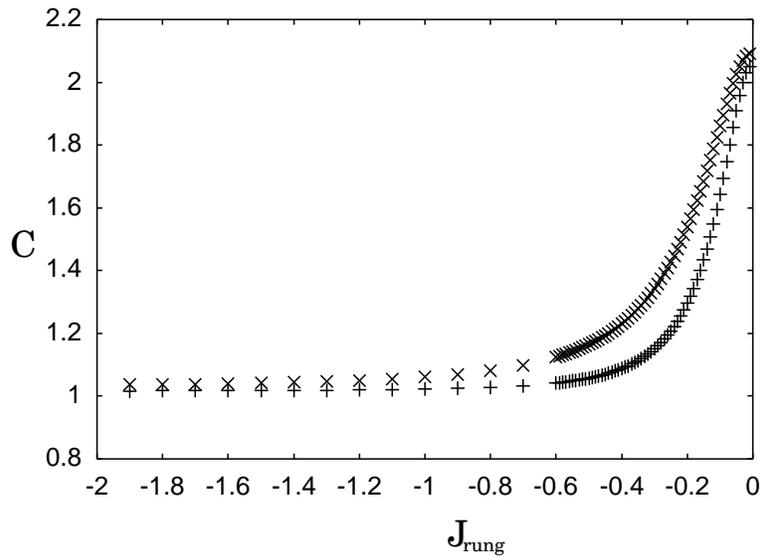,width=10cm}
\caption{The effective central charge on XY-Haldane transition line.
$\times$'s are extrapolated using $L(=N/2)=6,8,10$.
$+$'s are extrapolated using $L(=N/2)=8,10,12$.
}
\label{fig:cc-xy-haldane}
\end{center}
\end{figure}

\begin{figure}
\begin{center}
\epsfig{file=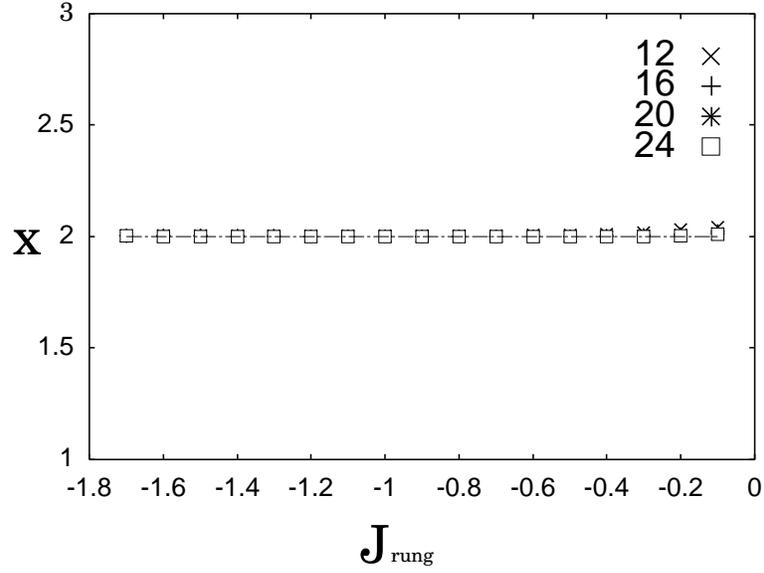,width=10cm}
\caption{Size dependence of scaling dimensions removed logarithmic
 correction on the XY-Haldane transition line ($\Delta=0$)
 for the system size $N(=2L)=12,16,20,24$.
}
\label{fig:sd-xy-haldane}
\end{center}
\end{figure}

The procedure in the previous subsection can be applied to the
$XY$-Haldane transition, since we can expect that $XY$-Haldane transition is
BKT type.
In this case, we can identify critical points as cross points of two
low-lying excitations. 
One has quantum number $S^z_T=\pm 2, P=1, q=0$ under the PBC,
another has quantum number $S^z_T=0, P^*=-1$ under the TBC.
We show crossing points in Fig.(\ref{fig:cp-xy-haldane}).
We can find the transition line is on $\Delta=0$.
As before, we calculate the effective central charge 
and the scaling dimensions after removing logarithmic correction.
We show results in Fig. \ref{fig:cc-xy-haldane} for the central charge, 
and in Fig. \ref{fig:sd-xy-haldane} for the scaling dimensions.
As a result, we obtain the phase boundary is on a $\Delta =0$ line.
This is analogous to the phase boundary 
between XY phase and Haldane phase in $S=1$ $XXZ$ spin chain
\cite{dN-R,C-H-S}.

We have found analytically that S=1 $XY$ spin chain has an additional 
$SU(2)$ symmetry under the OBC and an artificial boundary
condition(ABC) in our previous paper\cite{K-H-N}.
In this paper, we show that there is an additional $SU(2)$
symmetry for $S=1/2$ two-leg $XXZ$ spin ladder system with $\Delta=0$
under the OBC and ABC (see Appendix \ref{sec:addsu2} and Ref.\cite{K-H-N}).
This supports our numerical calculations,
since an $SU(2)$ symmetry is related to the BKT transition
\cite{Ginsparg}.

\subsection{Multicritical point $(\Delta,J_{\rm rung})=(0,0)$}
In this subsection, we investigate excitations near the multi-critical
point $J_{\rm rung}=0$ and $\Delta=0$.
On this point, BKT transition line meets Gaussian transition line.
BKT transition line is a phase boundary between $XY$ phase and Haldane
phase, and between $XY$ phase and rung-singlet phase.
Gaussian transition line is a phase boundary between rung-singlet phase
and Haldane phase.
BKT transition line consists of two parts,
one is the crossing line ($S^z_T=\pm 2, P=1, q=0$ under the PBC) 
and ($S^z_T=0, P^*=1$ under the TBC),
another is the crossing line ($S^z_T=\pm 2, P=1, q=0$ under the PBC)
and ($S^z_T=0, P^*=-1$ under the TBC).

\begin{figure}
\begin{center}
\epsfig{file=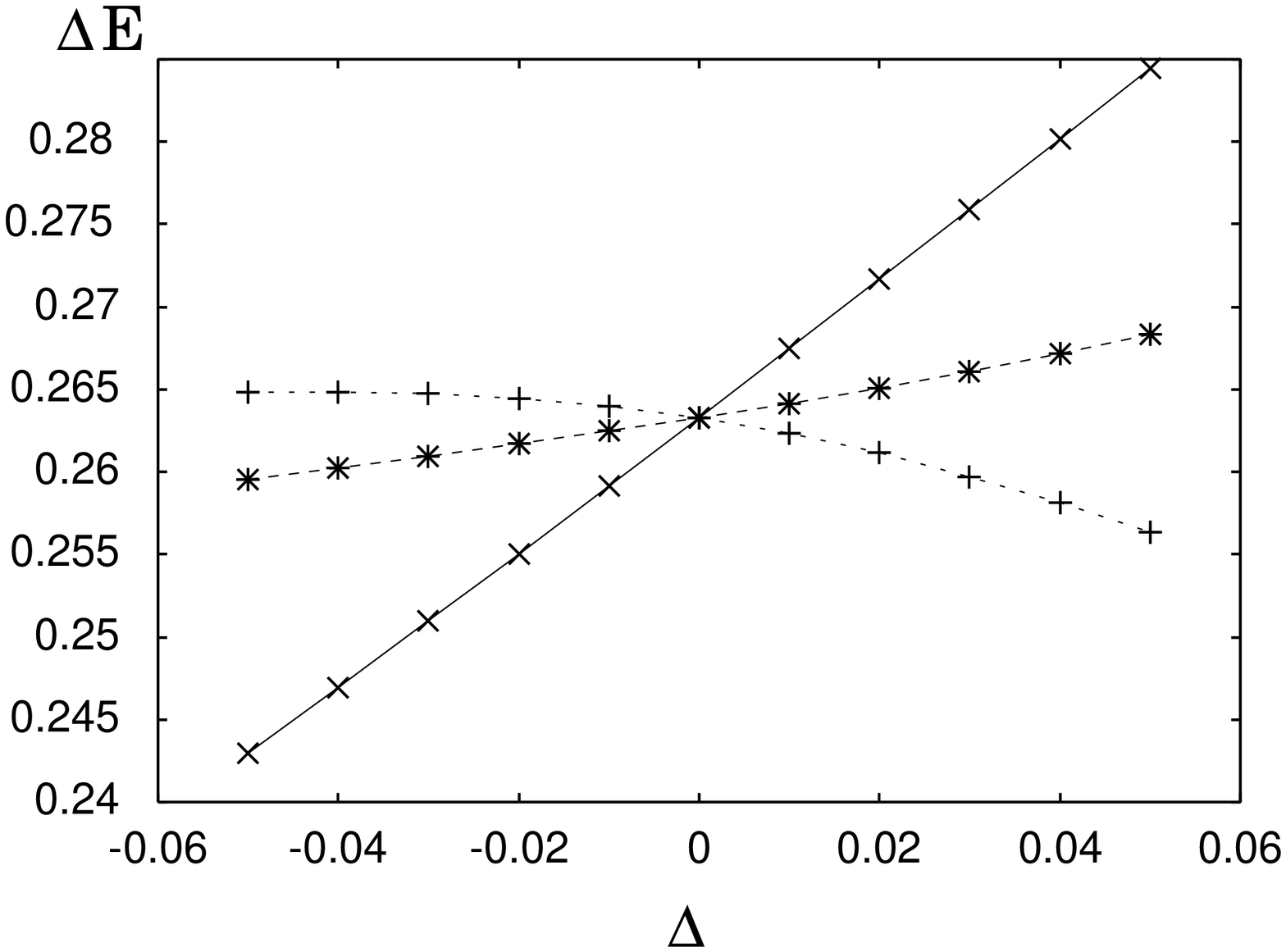,width=10cm}
\caption{
Excitation energies of $L=12(=N/2)$, on $\Delta=J_{\rm rung}$ line 
near the multicritical point. 
$\times$'s are $S^z_T=\pm2$,$q=0$,$P=1$ under
the periodic boundary condition. $+$'s are $S^z_T=0$,$P^*=1$ under
the twisted boundary condition. $*$'s are $S^z_T=0$,$P^*=-1$ under
the twisted boundary condition.
 }
\label{fig:mcp-xy-rs-o}
\end{center}
\end{figure}

\begin{figure}
\begin{center}
\epsfig{file=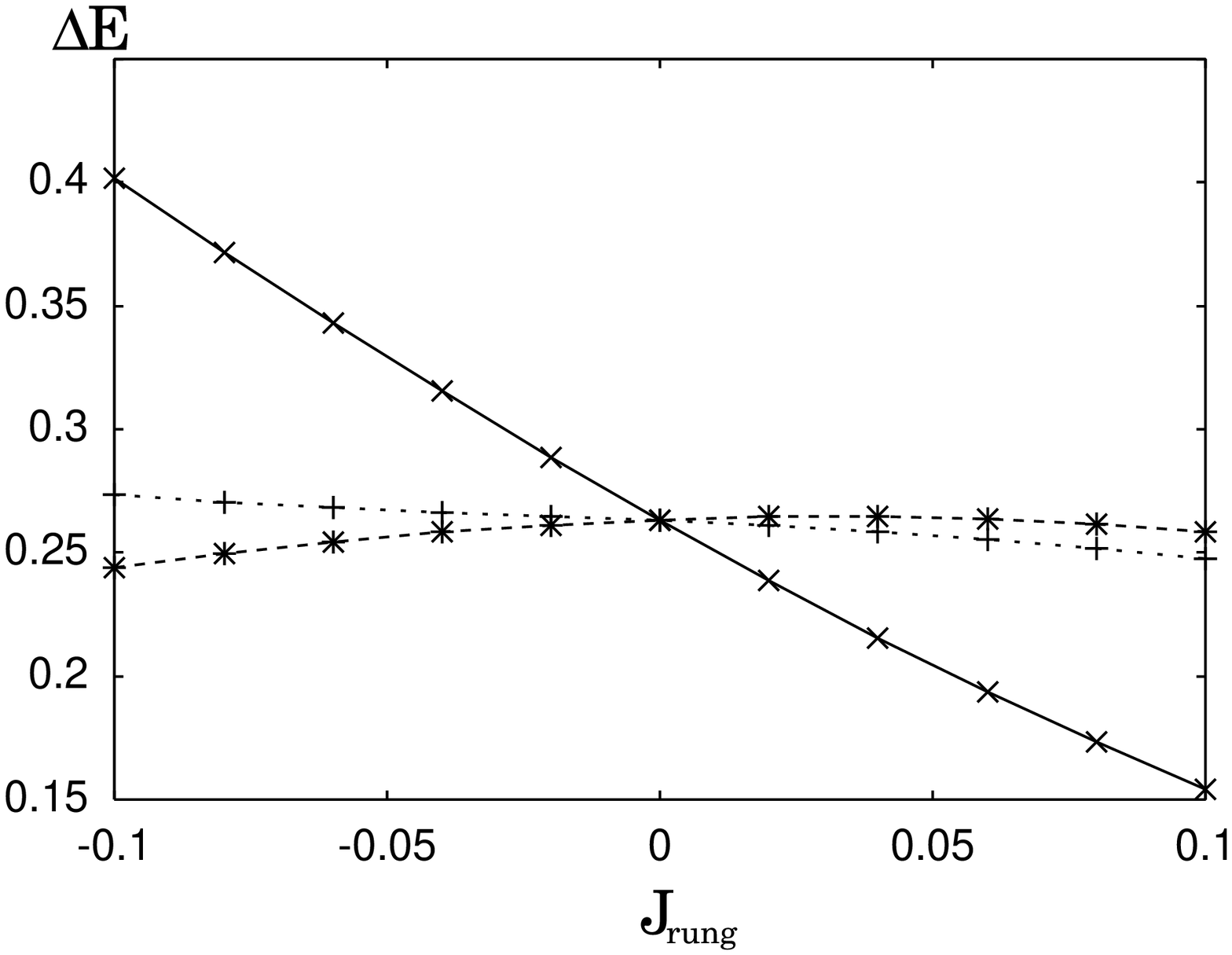,width=10cm}
\caption{
Excitation energies of $L=12(=N/2)$, on $\Delta=-4J_{\rm rung}$ line 
near the multicritical point. 
$\times$'s are $S^z_T=\pm2$,$q=0$,$P=1$ under
the periodic boundary condition. $+$'s are $S^z_T=0$,$P^*=1$ under
the twisted boundary condition. $*$'s are $S^z_T=0$,$P^*=-1$ under
the twisted boundary condition.
 }
\label{fig:mcp-xy-h-o}
\end{center}
\end{figure}

\begin{figure}
\begin{center}
\epsfig{file=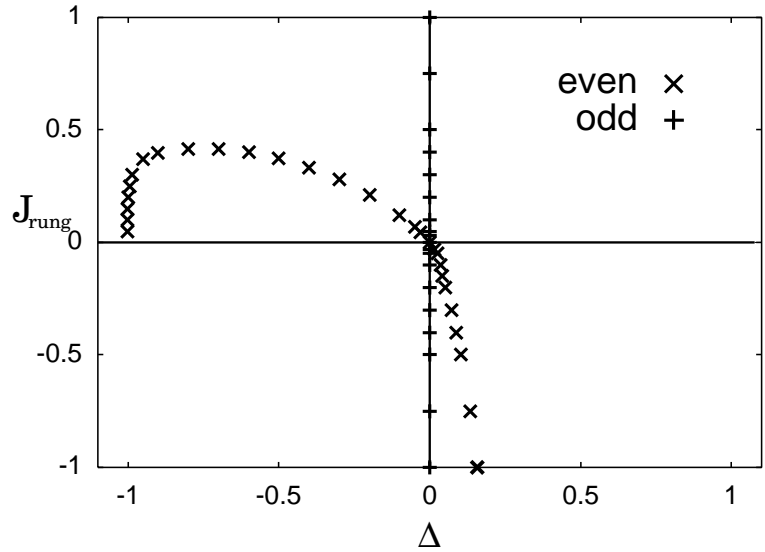,width=10cm}
\caption{
Extrapolated cross points of excitation energies.
$\times$'s are cross points of the excitation with
$S^z_T=\pm2$,$q=0$,$P=1$ under the periodic boundary condition and
the excitation with $S^z_T=0$,$P^*=1$ under the twisted boundary condition.
$+$'s are cross points of the excitation with $S^z_T=\pm2$,$q=0$,$P=1$ under
the periodic boundary condition and the excitation $S^z_T=0$,$P^*=-1$
under the twisted boundary condition.
}
\label{fig:cp-parity}
\end{center}
\end{figure}

We show excitation energies with the following quantum numbers and 
boundary conditions:
\begin{itemize}
 \item $S^z_T=\pm 2, P=1, q=0$ under the PBC
 \item $S^z_T=0, P^*=1$ under the TBC
 \item $S^z_T=0, P^*=-1$ under the TBC
\end{itemize}
in Fig. (\ref{fig:mcp-xy-rs-o}) along the $J_{\rm rung}=\Delta$ line
and in Fig. (\ref{fig:mcp-xy-h-o}) along the $J_{\rm rung}=-4\Delta$
line in the parameter space.
We show these crossing points in Fig.(\ref{fig:cp-parity}).
This show that $(\Delta,J_{\rm rung})=(0,0)$ is a multicritical point.
$XY1$-Haldane transition line and $XY2$-rung singlet transition line 
are continuously connected but these are not smoothly connected.

From bosonization studies \cite{S-M,V-J-M}, $XY$-rung singlet transition
line seems to smoothly connect $XY$-Haldane transition line.
However, this is not consistent with CFT analysis.
Our numerical calculations support CFT analysis.
We can consider that this reflects an additional SU(2) symmetry 
on $\Delta=0$ line \cite{K-H-N}.

\subsection{Multicritical point $(\Delta,J_{\rm rung})=(-1,0)$}
In this subsection, we discuss the other multicritical point 
$(\Delta,J_{\rm rung})=(-1,0)$.
This multicritical point is among the $XY$ phase, the Ferromagnetic
phase, the Stripe Ferromagnetic phase, and the rung-single phase.
This multicritical point has been studied by Kolezhuk and Mikeska \cite{K-M}
and Vekua {\it et al}.\cite{V-J-M}.
Kolezhuk and Mikeska discussed this,
mapping onto the anisotropic non-linear $\sigma$ model.
The renormalization group analysis of the anisotropic non-linear
$\sigma$ model was developed by Nelson and Pelcovits\cite{N-P}.
Vekua {\it et al}. also discussed the ferromagnetic leg case,
using the bosonization \cite{V-J-M}.
Outlines of the phase diagram are the same, 
however there is different point.
Kolezhuk and Mikeska suggested that the shape of the phase boundary
between $XY$ phase and rung-singlet phase is the exponential function
$J_{\rm rung} \propto  \exp \left(-a/\left|\Delta+1\right|\right)$,
where $a$ is a positive constant.
On the other hand, Vekua {\it et al}. suggested 
that the shape of phase boundary between the $XY$ phase and rung-singlet phase
is the linear line on $\Delta-J_{\rm rung}$ plane.
Our numerical result strongly supports Kolezhuk's suggestion
(See Fig.\ref{fig:pd-xy-rs}).
This shows that non-linear $\sigma$ model with anisotropy 
is more appropriate to describe the region near the multicritical point.

\section{CONCLUSIONS}
In this paper, we have studied the ground state phase diagram of  S=1/2
two-leg spin ladder system.
We have accurately numerically determined the phase boundary 
between the $XY$ phase and the Haldane phase,
and between the XY phase and the rung-singlet phase,
analyzing the exact diagonalization data using the level spectroscopy
method, TBC, and CFT.
And we have checked the universality class.
As a result, in $-1< \Delta < 0$ region, the $XY$ phase extends over the
$J_{\rm rung}=0$ line from the $J_{\rm rung}<0$ region to 
the $J_{\rm rung}>0$ region. 
And there does not exist the $XY$ phase in $\Delta>0$ region
except on the $J_{\rm rung}=0$ line, 
since the BKT transition line is the $\Delta=0$ line.
We can understand this result, considering an additional $SU(2)$
symmetry \cite{K-H-N}.

We have roughly discussed phase boundaries between the rung-singlet
phase and N\'eel phase, and between the rung-singlet phase and 
the stripe ferromagnetic phase, using the variational method.
We think that these phase transitions are the second order phase
transitions, considering the symmetry,
and that the universality class is the Ising type. 

We have determined phase boundaries between the stripe ferromagnetic
phase and the fully ferromagnetic phase, and between the N\'eel phase
and the stripe N\'eel phase, considering degeneracies of ground states. 
These phase transitions are of the first order type.

We find that there are two different $XY$ phase.
We can distinguish these $XY$ phases, considering $XX$ (or $YY$)correlation
function.
This is based on the Marshall-Lieb-Mattis's theorem.
One $XY$ phase ($XY2$) is the $J_{\rm rung}>0$ region, other $XY$ phase ($XY1$) is
the $J_{\rm rung}<0$ region,
and it is understood that the second order phase transition occurs
numerically between the $XY1$ phase and the XY2 phase
from Fig. \ref{fig:ecc} and Fig. \ref{fig:ratio}.

$J_{\rm rung}=0,\Delta=-1$ point is a multicritical point, among 
the rung-singlet phase, the stripe ferromagnetic phase,
the fully ferromagnetic phase, and the $XY$ phase.
Seeing the $J_{\rm rung}>0$ region, the phase diagram is quite characteristic.
This phase diagram is similar to one of the anisotropic 
nonlinear $\sigma$ model without the topological term
whose phase diagram and the renormalization group flow
have been discussed by Nelson and Pelcovits \cite{N-P}.
The same analysis can be applied to this problem.

\section*{ACKNOWLEDGEMENTS}

Numerical calculations in this paper was based on T.I.T.pack ver.2
coded by Prof. H. Nishimori.

\appendix

\section{Excitations of the two independent chains system}
\label{sec:ind-chain}
In $S=1/2$ two-leg ladder system,
the $J_{\rm rung}=0$ case is described as two independent $S=1/2$ XXZ chains.
An $S=1/2$ XXZ chain is exactly solved \cite{C-G}, 
using the Bethe ansatz \cite{Bethe}.
This model is described by the Hamiltonian
\begin{equation}
{\cal H}=J \sum_{i} \left( S^x_{i} S^x_{i+1} + S^y_{i} S^y_{i+1} 
+ \Delta S^z_{i} S^z_{i+1}  \right) .
\end{equation}
For $\Delta >1$, the system becomes the N\'eel (Anti Ferromagnetic)
phase, whereas for $\Delta <-1$, the system becomes the Ferromagnetic
phase. In the intermediate case $-1 \leq \Delta \leq 1$, the system
becomes XY (spin fluid) phase.

In this appendix, we discuss the excitation structure of two independent
S=1/2 XY spin chain.

In XY phase, the system is described as $c=1$ gaussian model.
Now we have two independent chains.
We describe the state of all systems as $\Psi$.
We think  that $\Psi_1$ is the state of chain 1,
and $\Psi_2$ is that of chain 2.
\begin{itemize}
 \item $\Psi = \Psi_1 \otimes \Psi_2$
 \item $q=q_1+q_2$ : wave number
 \item $S^T_z=S^{T1}_z+S^{T2}_z$ : magnetization
\end{itemize}

Excitations of the single chain are, for example,
\begin{itemize}
\item $q=0$,$S^T_z=0$,$x=0$ : the ground state
\item $q=\pi$,$S^T_z=0$,$x=1$
\item $q=0$,$S^T_z=0$,$x=2$
\item $q=\pi$,$S^T_z=\pm1$,$x=\frac{1}{4}$.
\end{itemize}

Excitation of the system are as follows,
\begin{itemize}

\item $ 
\left( q=\pi, S^{T1}_z=0, x=1 \right) 
\otimes
\left( q=\pi, S^{T2}_z=0, x=1 \right) 
=
\left( q=0(2\pi), S^{T}_z=0, x=2 \right) 
$

\item $ 
\left( q=0, S^{T1}_z=0, x=2 \right) 
\otimes
\left( q=0, S^{T2}_z=0, x=0 \right) 
=
\left( q=0(2\pi), S^{T}_z=0, x=2 \right) 
$

\item $
\left( q=\pi, S^{T1}_z=+1, x=\frac{1}{4} \right) 
\otimes
\left( q=\pi, S^{T2}_z=-1, x=\frac{1}{4} \right) 
=
\left( q=0(2\pi), S^{T}_z=0, x=\frac{1}{2} \right) 
$

\end{itemize}

The last one transforms as follows,
under the spin reversion $S_z \leftrightarrow -S_z$,

\begin{center}

$\left( q=\pi, S^{T}_z=+1, x=\frac{1}{4} \right)
\otimes 
\left( q=\pi, S^{T}_z=-1, x=\frac{1}{4} \right)$ \\
$\Downarrow$ \\
$ \left( q=\pi, S^{T}_z=-1, x=\frac{1}{4} \right)
\otimes 
\left( q=\pi, S^{T}_z=+1, x=\frac{1}{4} \right)  $
\end{center}

This has the odd rung-parity.
This excitation is relevant, it does not appear
in the space where the rung-parity is restricted to even.

\section{An additional $SU(2)$ symmetry of the 
$S=1/2$ two-leg spin ladder system\cite{K-H-N}}
\label{sec:addsu2}
We present a brief review of an additional $SU(2)$ symmetry 
of the $S=1/2$ two-leg ladder system.
In the $S=1$ XY chain and the $S=1/2$ two-leg ladder system,
there exists an additional $SU(2)$ symmetry 
besides the usual spin $SU(2)$ symmetry.
In our previous paper \cite{K-H-N},
we show that the one dimensional spin-$1$ XY model has
an additional $SU(2)$ symmetry.
And we also show the case of $S=1/2$ two-leg ladder system.

We consider the Hamiltonian (\ref{eq:Hamiltonian}).
At first, we introduce the following local operators:
\begin{equation}
\tilde{s}^{\pm}_j=S^{\pm}_{1,j}S^{\pm}_{2,j}
, \hspace{1cm} 
\tilde{s}^z_j=\frac{1}{2}\left(S^z_{1,j}+S^z_{2,j}\right).
\end{equation}
After the simple calculation,
we obtain the commutation relation of $SU(2)$,
\begin{equation}
\left[\tilde{s}^z_j, \tilde{s}^{\pm}_k \right]=\pm\delta_{jk}
\tilde{s}^{\pm}_j , \hspace{1cm}
\left[\tilde{s}^+_j, \tilde{s}^-_k\right]=2\delta_{jk}\tilde{s}^z_j
\end{equation}
The operator $\sum_j \tilde{s}^z_j$ commutes with the Hamiltonian 
(\ref{eq:Hamiltonian}), but operators $\sum_j\tilde{s}^{\pm}_j$ do not.

Thus we introduce new non-local operators,
\begin{equation}
s^{\pm}_j=S^{\pm}_{1,j}S^{\pm}_{2,j}\prod^{j-1}_{l=1}
\left(-4S^z_{1,l}S^z_{2,l}\right) \ \ \ \
s^z_j=\frac{1}{2}\left(S^z_{1,j}+S^z_{2,j}\right).
\end{equation}
New operators satisfy the commutation relation of $SU(2)$,
\begin{equation}
\left[s^z_j,s^{\pm}_k\right]=\pm\delta_{jk}s^{\pm}_j  \ \ \ \
\left[s^+_j,s^-_k\right]=2\delta_{jk}s^z_j
\end{equation}
and
\begin{equation}
\left[s^z_T,s^{\pm}_T\right]=\pm s^{\pm}_T  \ \ \ \
\left[s^+_T,s^-_T\right]=2s^z_T.
\end{equation}
where $s^{\pm}_T=\sum_j s^{\pm}_j$ and 
$s^z_T\sum_j s^z_j$ are total operators.

After the some calculation,
we can show that these total operators
commute with the Hamiltonian (\ref{eq:Hamiltonian})
with $\Delta=0$ under the open boundary condition 
and an artificial boundary condition.
For a discussion in detail, please see our previous paper\cite{K-H-N}.

\newpage

\end{document}